\documentclass{pnastwo}

\usepackage{amssymb,amsfonts,amsmath}
\usepackage{latexsym}
\usepackage{graphicx}
\usepackage{rotating}

\usepackage{color}

\newcommand{\dz}{d$_{3z^{2}-r^{2}}$ }
\newcommand{\dx}{d$_{x^{2}-y^{2}}$ }
\newcommand{\dxy}{d$_{xy}$ }

\newcommand{\dzd}{d$_{3z^{2}-r^{2}}$}

\newcommand{\dxyd}{d$_{xy}$}

\newcommand{\oxy}{O$_2$ }
\newcommand{\oxyd}{O$_2$}

\newcommand{\emphasize}{\emph}

\def\onlinecite#1{\cite{#1}}
\newcommand{\up}{\uparrow}
\newcommand{\dn}{\downarrow}

\usepackage[compact]{titlesec}
\titlespacing{\section}{0pt}{*0}{*0}
\titlespacing{\subsection}{0pt}{*0}{*0}
\titlespacing{\subsubsection}{0pt}{*0}{*0}
\contributor{Accepted for publication in the Proceedings
of the National Academy of Sciences of the United States of America (2014).}
\url{www.pnas.org/cgi/doi/10.1073/pnas.1322966111}
\copyrightyear{2014}
\issuedate{Issue Date}
\volume{Volume}
\issuenumber{Issue Number}

\begin{document}


\title{Renormalization of myoglobin-ligand binding energetics by quantum many-body effects}

 \author{C\'edric Weber\affil{1}{Theory and Simulation of Condensed Matter, KingÕs College London, London WC2R 2LS, United Kingdom}\affil{2}{Thomas Young Centre, University College London, London WC1H 0AH, United Kingdom}, Daniel J. Cole\affil{3}{Department of Chemistry, Yale University, New Haven, CT 06520-8107}\affil{4}{Cavendish Laboratory, University of Cambridge, Cambridge CB3 0HE, United Kingdom}, David D. O'Regan\affil{5}{School of Physics and the Centre for Research on Adaptive Nanostructures and Nanodevices (CRANN), Trinity College Dublin, Dublin 2, Ireland}\affil{6}{Theory and Simulation of Materials, \'Ecole Polytechnique F\'ed\'erale de Lausanne, 1015 Lausanne, Switzerland}, \and Mike C. Payne\affil{4}{}}

\maketitle


\begin{article}

\begin{abstract} 
  We carry out a first-principles atomistic study of the electronic
  mechanisms of ligand binding and discrimination 
  in the myoglobin protein.
 Electronic correlation effects are taken into account using one of
 the most advanced methods currently available, 
 namely a linear-scaling density
 functional theory (DFT) approach wherein the treatment of localized
 iron $3d$ electrons is further refined using 
 dynamical mean-field theory (DMFT).
 This combination of 
 methods explicitly accounts for dynamical and multi-reference
 quantum physics, such as valence and spin fluctuations, of the $3d$
 electrons, whilst treating a significant proportion of the protein
 (more than 1000 atoms) with density functional theory.
 The computed electronic structure of the myoglobin complexes and the
 nature of the Fe--\oxy bonding are validated against experimental
 spectroscopic observables.
  We  elucidate and solve a long standing problem related to
  the quantum-mechanical description of the respiration process,
  namely that DFT calculations predict a strong imbalance between
  O$_2$ and CO binding, favoring the latter to an unphysically large
  extent.
  We show that the explicit inclusion of many body-effects induced by
  the Hund's coupling mechanism results in the correct prediction of
  similar binding energies for oxy- and carbonmonoxymyoglobin.

\end{abstract}
%

\begin{section}{Significance Statement}

Heme-based metalloproteins play a central role in respiration by transporting and storing oxygen, a function that is inhibited by carbon monoxide. Density-functional theory has been unable to provide a complete description of the binding of these ligands to hemeÕs central iron atom, predicting an unrealistically high relative affinity for carbon monoxide. Here, we solve this problem using dynamical mean-field theory in combination with linear-scaling density-functional theory, thus allowing for a simultaneous description of crucial quantum entanglement and protein discrimination effects in the ground-state of the oxygen-heme complex. By simulating the binding process within a 1,000-atom quantum-mechanical model of the myoglobin metalloprotein, we obtain a significantly improved description of its spectroscopic and energetic observables.

\end{section}
\section{Introduction}
\dropcap{T}he ability of metalloporphyrins to bind small
ligands is of great interest in the field of biochemistry.
One such example is the heme molecule, which
reversibly binds diatomic ligands,
such as oxygen (\oxyd) and carbon monoxide (CO), and plays a crucial
role in human respiration.
Heme is employed in myoglobin (Mb) and hemoglobin (Hb) proteins to
 store and transport \oxy in vertebrates.
The heme group of Mb is packed within a predominantly $\alpha$-helical
secondary structure and is coordinated by a histidine residue (known
as the proximal histidine) as the fifth ligand of the heme's central
Fe ion.

Despite intensive
studies~\cite{pauling64,weiss64,goddard75,chen08,ribas07,radon08},
the nature of the bonding of \oxy  to the iron
binding site of the heme molecule remains poorly understood, mainly due
to the strong electronic correlation effects associated with its  
localized Fe $3d$ electrons. It is known that these electrons are
energetically well-aligned with the $\pi^*$ acceptor orbitals of CO and
\oxyd, and that the molecules' bound conformations seek to maximize 
intermolecular orbital overlap~\cite{reed77,review,vojtechovsky99}.
In the case of Mb\oxyd, the short Fe--O bond
(1.81~\AA{}~\cite{vojtechovsky99}) implies that $\sigma$-bonding is
supplemented, to some extent, by $\pi$-bonding~\cite{review}.
Indeed, calculations employing the {\it ab initio} complete active
space self-consistent method, in combination with a molecular
mechanics force field to describe the protein (CASSCF/MM)~\cite{chen08}, have
identified a weak $\pi$-bonding mechanism in the Fe--\oxy bond that
gives rise to an antiferromagnetic (open-shell singlet) state.

However, recent Fe L-edge X-ray absorption spectroscopy measurements
on small biomimetic heme models lack the signature low-energy peak that
is characteristic of the $d\pi$ hole, formed by metal-to-ligand charge
transfer into the ligand $\pi^*$ orbitals~\cite{wilson13}.
Although these spectroscopic results are more consistent with a strong
Fe--O $\pi$ interaction, some uncertainty remains about whether the same
bonding picture holds in Mb\oxyd, since the experiment was performed
on a small model system (Fe(pfp)1-MeIm\oxyd), which, in particular,
neglects the distal histidine (His\,64) that hydrogen bonds directly
with \oxy in the protein.

Furthermore, while the diamagnetic nature of Mb\oxy is
well-established, there is little experimental evidence that directly 
addresses the extent of the charge transfer from Fe to \oxyd.
Both CASSCF/MM~\cite{chen08} and L-edge X-ray absorption
spectroscopy~\cite{wilson13} suggest strong $\sigma$-donation from
\oxy into the \dz orbital of iron (ligand-to-metal back charge
transfer), which is hypothesized to limit the charge on the \oxy
molecule to around $-0.5$~e~\cite{chen08}.
However, the stretching frequency of the O--O bond in Mb\oxy has been
shown to be close to that of the free O$_2^-$
ion~\cite{review2,review}, which motivates further study.

The energetics of diatomic ligand binding to the Mb protein are
expected to depend strongly on the electronic structure of the heme
site, and in particular on its orbital polarization.
Specifically, Mb reduces the heme group's natural preference for CO
binding: based on experimental equilibrium association constants, the
binding free energy of CO, relative to \oxyd, is reduced from around
5.9~kcal/mol in a non-polar solvent to 1.9~kcal/mol in the protein
environment~\cite{olson97}.
The combination of the strong electronic correlation centered at 
the Fe binding site and
long-ranged interactions between the protein and the charged \oxy
molecule make computational modeling of the energetics of these
complexes extremely challenging.

Typically, such studies calculate the protein
effect~\cite{cole,sigfridsson02,angelis05}, or relative spin state
energies~\cite{heme_marzari}, or focus on small model
systems~\cite{radon08}. However, most approaches applied to
large system sizes did not include a proper treatment of electronic correlations.
The effect of electronic correlations in the iron $3d$ states was investigated by some of us~\cite{cole}, 
where it was included, to an extent, in {\it ab initio} simulations of ligand discrimination in 
myoglobin via a DFT+$U$ treatment.
DFT+$U$ has been shown to be an efficient method
for correcting self-interaction errors in the approximate DFT description of
transition-metal chemistry~\cite{PhysRevLett.97.103001}, and when combined
with linear-scaling approaches~\cite{dftu_david_subspace_representation,linearscalingdftu}, it allows us to tackle
such systems comprising thousands of atoms.
It was found, in the case of Mb~\cite{cole}, 
that the protein discrimination effect is dominated by
polar interactions between \oxy and the distal protein residue His\,64.
However, a problem in the DFT+$U$ calculations is that a strong residual energetic imbalance that favors CO over
\oxy binding was observed~\cite{cole}, suggesting that approaches beyond static DFT+$U$ are called for in order to
obtain a proper description of myoglobin.

Recent progress has been made in the study of strongly-correlated
electrons by means of dynamical mean-field theory
(DMFT)~\cite{OLD_GABIS_REVIEW}, a sophisticated method that includes
quantum dynamical effects, and takes into account both valence and
spin fluctuations. DMFT is routinely used to describe materials, and recently
has also been extended to nanoscopic systems \cite{referee1,referee2}.
DMFT also explicitly includes the Hund's exchange coupling typically,
although not always, neglected in DFT and DFT+$U$ studies.
DMFT was recently combined with linear-scaling DFT~\cite{onetep} to
produce a linear-scaling DFT+DMFT approach~\cite{our_paper_vo2}.
By means of the latter, we have pointed out that strong correlation
effects in heme are controlled by the Hund's coupling $J$, and not the
Hubbard repulsion $U$ alone~\cite{weber13}, suggesting that subtle
quantum many body effects are missing in the DFT+$U$ treatment of
myoglobin~\cite{cole}.
However, the computational model included just the heme group and
diatomic ligands, neglecting entirely the protein environment and proximal histidine ligand,
thereby strongly overestimating the binding energy of CO relative to O$_2$.

In the present work, we bridge state-of-the-art DMFT many-body calculations
with large scale DFT calculations. We perform simulations of realistic models of the
Mb\oxy and MbCO complexes, comprising 1007 atoms, 
using linear-scaling DFT+DMFT. We thereby 
treat the electrostatic, steric, and
hydrogen-bonding effects due to protein materials, together with the
multi-reference, finite-temperature and explicit Hund's exchange
coupling effects associated with the iron $3d$ binding site, 
in single, self-consistent calculations for the first time.
We systematically investigate how the Hund's coupling alters the
electronic structure at the heme site and, at the same time, corrects
the long-simulated, unphysical imbalance between CO and \oxy binding
affinities.


\section{Results}

\begin{figure}
\begin{center}
\includegraphics[width=0.8\columnwidth]{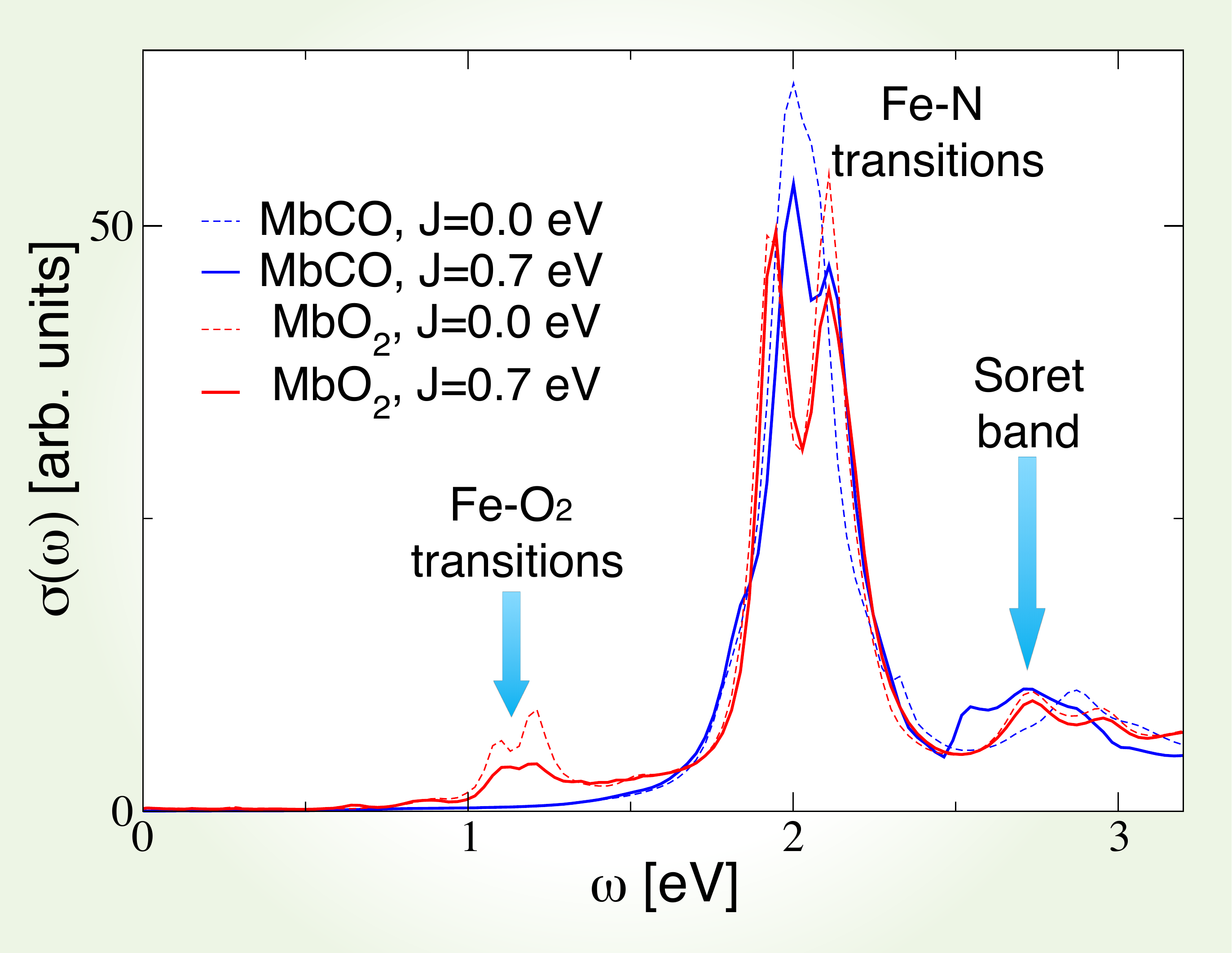}
\caption{ {\bf Optics:} porphyrin-plane component of the
optical absorption spectrum calculated using DFT+DMFT, at
  $J=0$~eV (dashed curves) and $J=0.7$~eV (continuous curves), for MbCO
  (blue) and MbO$_2$ (red).}
\label{fig_optic}
\end{center}
\end{figure}

We first discuss our results for the porphyrin-plane component
of the optical absorption spectra of ligated myoglobin, computed using
DFT+DMFT (Fig.~\ref{fig_optic}), where a realistic value of the Hund's coupling
$J=0.7$~eV is considered (results obtained for $J=0$~eV are shown for comparison).
The absorption spectrum of this protein has been reported experimentally
to be qualitatively dependent on its ligation state, in that a peak is present
in the infrared region in the MbO$_2$ case but not 
for MbCO~\cite{nozawa76}.
Our theoretical absorption spectrum is in good
agreement with the experimental data obtained from sperm whale MbO$_2$
single crystals~\cite{optics_sperm_whale}, and reproduces an
MbO$_2$ infrared absorption band at $\approx 1.2$~eV, 
observed experimentally at $1.3$~eV~\cite{nir-1}.
Our calculations associate this feature with a charge transfer band
generated by the hybridization of the Fe atom and the O$_2$ molecule,
in particular transitions of occupied porphyrin $\pi$ and iron
$d$ orbital states into empty O$_2$ ($\pi^*$) orbitals.
In MbCO, due to the strong covalent bond, the porphyrin $\pi$ and $d$
hybridized orbitals are at a lower energy and, hence, there is no
contribution to the infrared spectrum.
The double peak structure in the optical transition obtained at
$\omega \approx 1.9$~eV and 2.2~eV is also very close to experiments,
where they are obtained respectively at 2.1 and
2.3~eV~\cite{nir-1,optics_sperm_whale}.
We attribute this feature to the porphyrin Q band ($\pi$ to $\pi^\ast$
absorptions) and to corresponding charge transfer excitations.
We find a broad Soret band centered theoretically
at $\approx 2.75$~eV, close to the experimental peak obtained at 
2.95~eV~\cite{nir-1,optics_sperm_whale}.
For MbO$_2$, the spectrum at $J=0$~eV is qualitatively similar. 
However, Fig.~\ref{fig_optic} also reveals that a non-zero $J$ is required to recover the experimentally 
observed double-peak structure of the MbCO Q band~\cite{optics_sperm_whale}.
Analysis of the spectral weight below the Fermi level in MbCO reveals
the source of this splitting. For $J=0$~eV, the orbital character of
the HOMO is almost degenerate between the three $t_{2g}$
orbitals. However, for $J=0.7$~eV, we observe a splitting of the spectral weight of the d$_{xy}$ and the d$_{xz,yz}$ 
orbitals of $\approx 0.3$~eV, 
thus recovering the expected splitting of the charge transfer Q band in the optical absorption spectrum.

\begin{figure*}
\begin{center}
\includegraphics[width=2\columnwidth]{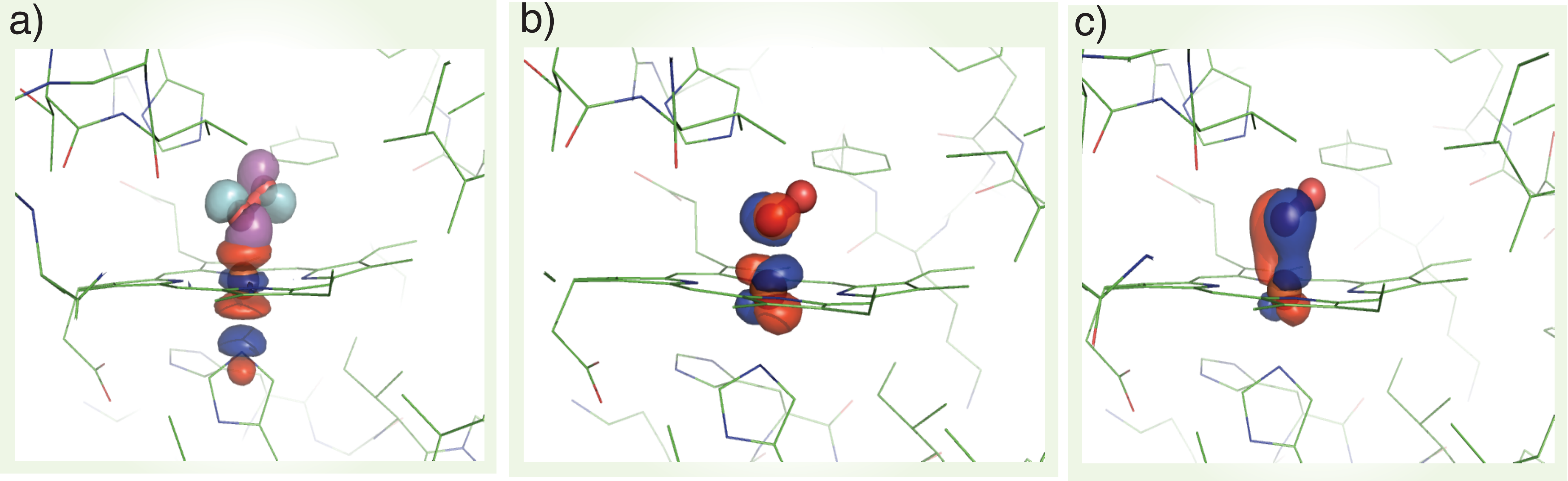}
\caption{{\bf Natural bond orbital (NBO)
analysis of the DFT+DMFT Green's function for Mb\oxyd:} 
(a) An \oxy $\pi^*$ type many-body NBO (cyan/magenta), 
of occupancy $1.5$~e, which strongly-interacts (via the DFT
Hamiltonian) with 
an anti-bonding Fe-based many-body NBO (red/blue), 
of occupancy $0.5$~e. 
(b,c) many-body NBOs showing $\pi$-bonding between
  Fe and \oxyd; the NBO occupancies are $1.4$~e and $1.8$~e, respectively.}
  \label{fig_orbs}
\end{center}
\end{figure*}

In order to further understand the nature of the bonding in the Mb\oxy
complex, we have found it instructive to transform the atomic basis
functions, used to expand the DFT+DMFT density-matrix (that is, the
frequency-integrated Green's function), into a set of natural bond
orbitals (NBOs)~\cite{reed88,nbo5,lee13}.
The transformation is constructed such that the resulting orbitals may
be categorized into localized Lewis-type bonding and lone pair
orbitals, as well as their anti-bonding and Rydberg counterparts, thus
allowing a chemically-intuitive population analysis to be applied to
the DFT+DMFT many-body wave-function.
Fig.~\ref{fig_orbs} shows the $\sigma$- and $\pi$-bonding many-body
natural bond orbitals of the Mb\oxy complex.
We find, in particular, that an \oxy $\pi^*$ NBO
(Fig.~\ref{fig_orbs}\textbf{a}), which has an occupancy of $1.5$~e,
and an anti-bonding NBO formed between Fe and the proximal histidine,
with occupancy $0.5$~e and a strong \dz character, interact strongly
via the DFT Hamiltonian.
We note that the $0.5$~e occupancy of the anti-bonding orbital on Fe
is consistent with the ligand--metal back charge transfer process
between \oxy and the Fe \dz orbital, which is observed both in
CASSCF/MM~\cite{chen08}, and with L-edge X-ray absorption
spectroscopy~\cite{wilson13}.
Ligand--metal back charge transfer is also present at $J=0$~eV, albeit
with a smaller magnitude ($0.34$~e).
Thus, electronic delocalization is expected to provide a greater
energetic stabilization in the Mb\oxy complex at $J=0.7$~eV.

The net charge on the \oxy molecule, from natural population analysis,
is $-1.1$~e, which is consistent with the Weiss picture of bonding in
Mb\oxyd~\cite{weiss64}.
It is worth noting that state-of-the-art CASSCF/MM calculations point
toward a smaller O$_2$ charge of $-0.5$~e~\cite{chen08}.
Metal-to-ligand charge transfer is expected to occur {\it via}
$\pi$-bonding interactions between Fe $d$ orbitals and \oxy
$\pi^*$~\cite{chen08,wilson13}.
Indeed, Fig.~\ref{fig_orbs}\textbf{b,c} are characteristic of the
multi-configurational CASSCF orbitals that make up the proposed
$\pi$-type bonding in a previous study~\cite{chen08}.
A notable difference between these calculations and the CASSCF/MM
study is that the $\pi$-bonding is much stronger than previously
reported.
Here, $3.25$~e are involved in $\pi$-bonding, as opposed to
approximately $2$~e in CASSCF/MM.
Our calculations yield a $d\pi$ hole character of 19~\%. This compares
extremely favorably with recent Fe L-edge X-ray absorption
spectroscopy measurements of a small biomimetic heme model, which 
estimates the $d\pi$ hole character to be $15\pm5$~\%~\cite{wilson13}.
We, therefore, find that the $\pi$-bonding character in Mb\oxy is
similar to that in isolated porphyrins.
We note that the stronger $\pi$-bonding interaction between the iron
and O$_2$ also suggests that spin polarization of the $\pi$ electrons
is less likely~\cite{wilson13}, suggesting that a broken spin symmetry
description of Mb\oxy might not be entirely reliable.

\begin{table}
  \caption{$3d$ orbital occupations of the Fe atom in ligated Mb. The CASSCF/MM values are shown
  for comparison \cite{chen08}.}
\begin{tabular} {c c c c c c }
 \hline
 Protein    &   d$_{x^2-y^2}$  & d$_{3z^2-r^2}$ & d$_{xy}$  & d$_{xz}$ &  d$_{yz}$       \\
 \hline
 MbO$_2$ ($J=0.0$~eV)    &  0.82  &  0.38  & 1.99 & 1.90 &  1.92    \\ 
 MbO$_2$ ($J=0.7$~eV)    &  0.96  &  0.96  & 1.10 & 1.87 &  1.96     \\
 \hline
 MbCO ($J=0.0$~eV)         &  0.95  & 1.13   & 1.99 & 1.88 &  1.89     \\
 MbCO ($J=0.7$~eV)         &  1.00  & 1.15   & 1.99 & 1.84 &  1.84     \\
 \hline
 Mb\oxyd (CASSCF/MM)     & 0.44   & 0.59   & 1.93 & 1.86 & 1.13      \\
\hline
\end{tabular}
\label{table}
\end{table}

\begin{figure}[b]
\begin{center}
\includegraphics[width=\columnwidth]{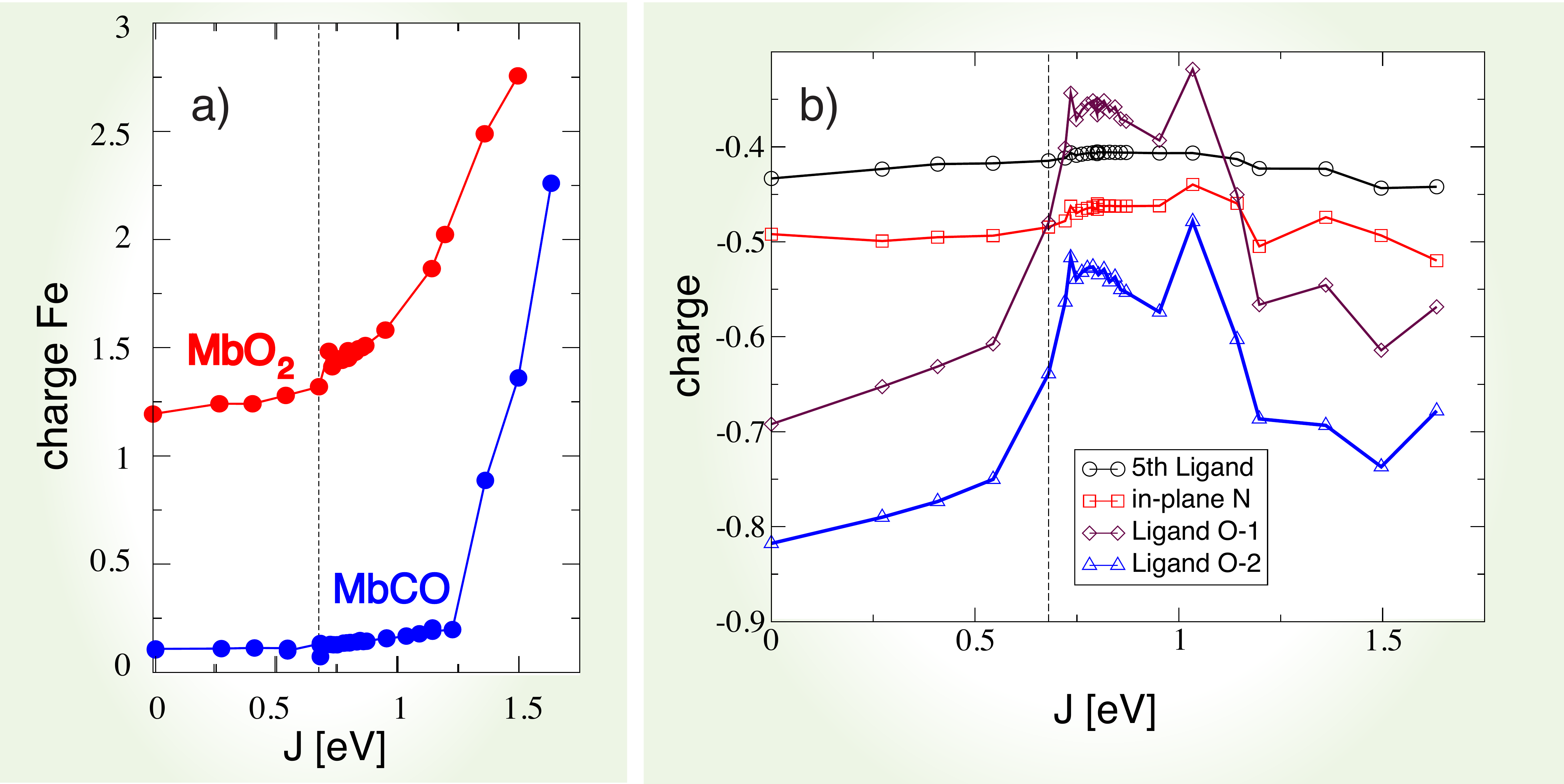}
\caption{ {\bf Polarization:}  
a) The dependence on $J$ of the DFT+DMFT Mulliken charge of the Fe atom in MbO$_2$ and MbCO. 
b) Mulliken charges in MbO$_2$ of the two O atoms of O$_2$ (O-1 is bonded to Fe and O-2 forms a hydrogen bond with His 64). Also shown are the corresponding charges of the in-plane N atoms of the heme group and His93 (the 5th ligand of Fe).}
\label{fig_charge}
\end{center}
\end{figure}

Next, we show, in Table~\ref{table}, a comparison of the 
computed Fe orbital
density with and without the explicit  inclusion of
Hund's coupling $J$ in the Hamiltonian.
We find that the effect of $J$ in MbO$_2$ is to
bring the \dzd, \dxyd, and \dx orbitals closer to single-electron
occupation, so that the Hund's coupling enhances the spin magnetic
moment on the Fe atom.
Indeed, we find, in our calculations, a build up of a magnetic moment in the
d$_{x^2-y^2}$, d$_{3z^2-r^2}$ and \dxy orbitals, with a concomitant
electron occupation of $n \approx 1$ absent from our $J=0$~eV calculation
and from the CASSCF/MM approach.
The latter discrepancy may be due to the fact that 
CASSCF does not include dynamical correlation effects 
and may also be dependent on the chosen active space.
In MbCO, unlike MbO$_2$, 
we observe that the doublet on the \dxy orbital is not emptied as the 
Hund's coupling is increased. 
We next show, in Fig.~\ref{fig_charge}\textbf{a}, the dependence of the Fe
charge, computed using Mulliken analysis in the NGWF basis,  
on the Hund's coupling $J$ for MbO$_2$ and MbCO, respectively.
For MbO$_2$, we find that the charge of the Fe is  
transferred to the porphyrin ring and protein as the Hund's coupling
is increased.
In contrast, for MbCO we find a very weak dependence of the Fe charge
on the Hund's coupling parameter (see Table~\ref{table}).
In our view, the latter indicates a very strong Fe-CO covalent
bond, which remains stable against the Hund's coupling.
We find that the charge transferred from the Fe to O$_2$
is $1.1$~e at the physical value of the Hund's coupling $J=0.7$~eV (see
Fig.~\ref{fig_charge}\textbf{b}), confirming the estimation obtained
using natural population analysis.

\begin{figure}[t]
\begin{center}
\includegraphics[width=0.8\columnwidth]{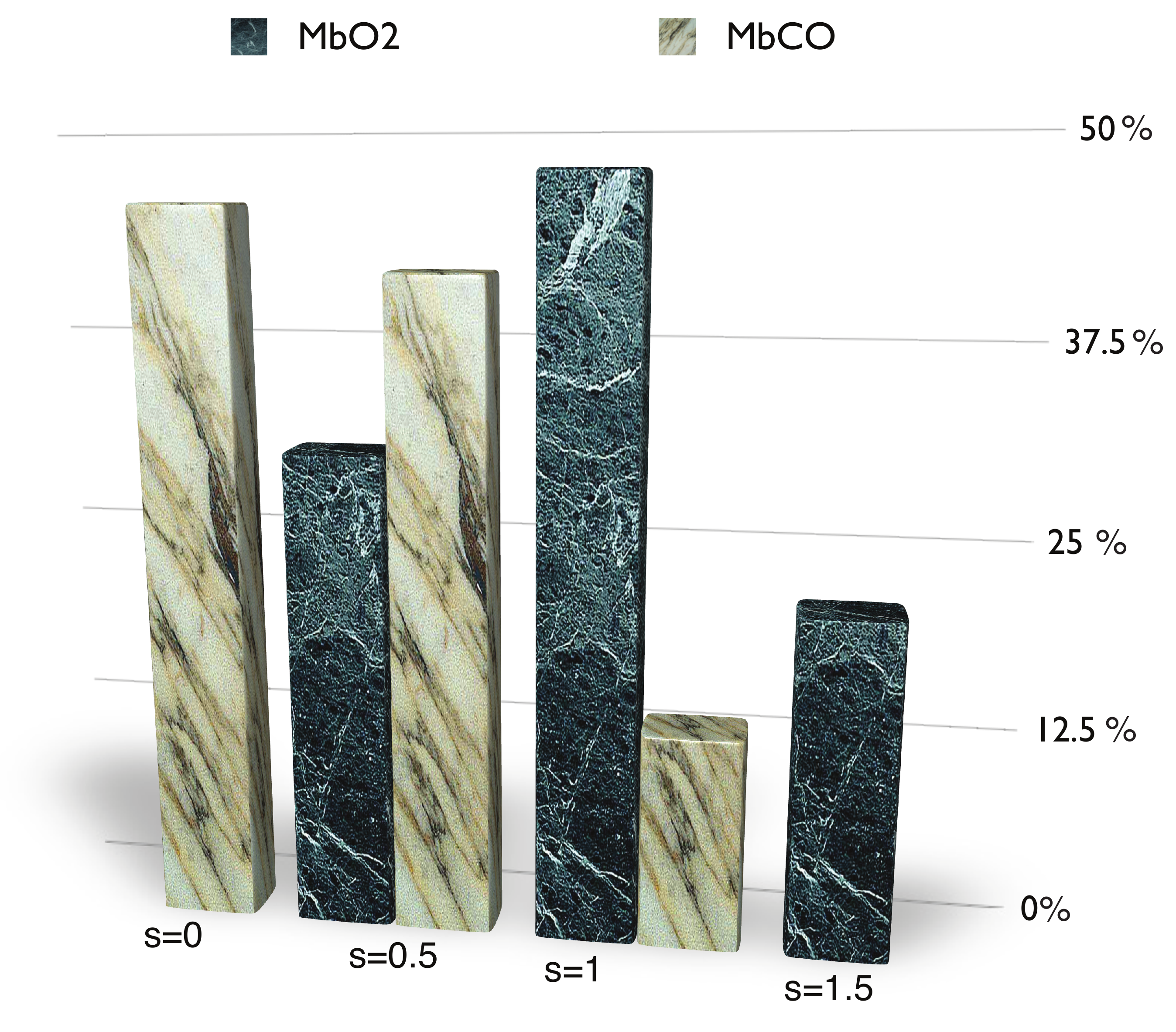}
\caption{ 
{\bf Spin fluctuations:}
Spin-state distribution
obtained by analyzing the 16 dominant 
states in the DFT+DMFT reduced density-matrix (see text) at $J=0.7$~eV.
The reduced density-matrix of MbCO has states with dominant $s=0$ configurations, and MbO$_2$
has dominant contributions from $s=1$, with higher spin contributions at $s=1.5$.}
\label{fig_spin}
\end{center}
\end{figure}

Fig.~\ref{fig_spin} depicts the computed 
spin fluctuations in MbO$_2$ and MbCO,
specifically, 
a histogram of the spin quantum number distribution obtained by looking 
at the 16 dominant states in the reduced ($3d$ subspace) density-matrix, obtained
by tracing the atomic DMFT problem over the bath degrees of freedom. This gives an effective representation of the quantum states of
the Fe atom. The ground-state wave-function is not a 
pure state with a single allowed value for the magnetic 
moment (singlet, doublet, triplet, etc.), yet we can
describe the fluctuating magnetic moment of the Fe atom by analyzing the distribution of the magnetic moments obtained from the dominant
configurations. In particular, we find that the reduced density-matrix of MbCO has states with dominant $s=0$ 
configurations, and MbO$_2$ has dominant contributions from $s=1$, with higher spin contributions at $s=1.5$.
This is consistent with our general 
observation that MbO$_2$ has larger valence fluctuations
(entanglement in the ground-state) than MbCO.

\begin{figure}
\begin{center}
\includegraphics[width=0.8\columnwidth]{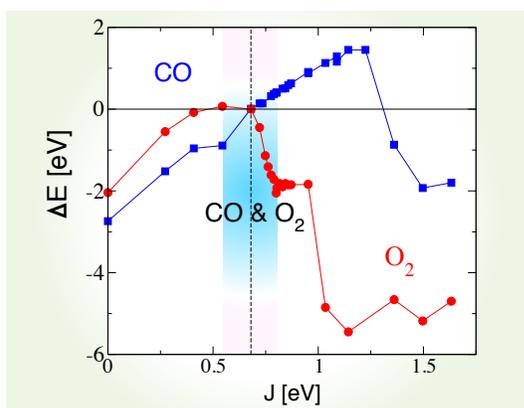}
\caption{ 
{\bf Binding energetics:} Hund's exchange coupling $J$-dependence of
the DFT+DMFT total-energy 
of MbCO (blue) and MbO$_2$ (red), where
the DFT energy of the diatomic ligand, CO or O$_2$, has been 
subtracted to give $\Delta E$. 
A global arbitrary constant shift of the energy has been applied for clarity.
The difference of binding energies $\Delta \Delta E$ is obtained by
$\Delta \Delta E = \Delta E (O_2) - \Delta E (CO) $. 
The physical regime, where $\Delta \Delta E$ is small, is shaded.}
\label{fig_energy}
\end{center}
\end{figure}

Having shown how the Hund's coupling affects the orbital
occupancy of the Fe site in Mb, and the associated charge transfer to
the \oxy molecule, we investigate, in what follows, 
how the energetics of ligand binding
to Mb are determined by these effects, and how the protein uses quantum
fluctuations to discriminate between \oxy and CO.
Fig.~\ref{fig_energy} shows the binding energies of \oxy and CO to Mb
(to within a constant shift), calculated using DFT+DMFT, as a function
of the Hund's exchange coupling $J$.
At $J=0$~eV, the binding energy of MbCO is approximately $1$~eV more favorable
than the binding energy of MbO$_2$, yielding an unphysical energetic
imbalance.
For $J>0.7$~eV, we find, on the contrary, that the binding energy of
MbO$_2$ is dramatically reduced.
We attribute this to the enhancement of the spin magnetic moment in
the Fe atom.
In the intermediate regime, close to $J=0.7$~eV, we find that the
imbalance between MbO$_2$ and MbCO is thereby also dramatically reduced.
In fact, the experimental binding free energy difference, between the
two ligands, of $1.9$~kcal/mol, is recovered from our calculations when
$J$ is near $0.7$~eV, a typical value used for iron-based materials~\cite{iron_arsenide}.
In this case, the effect of $J$ on the binding energy of \oxy may be
regarded as a balance between two competing effects.
The charge analysis (Fig.\ref{fig_charge}) reveals that
metal-to-ligand charge transfer is higher for $J=0$~eV, which is
expected to enhance ligand--protein interactions for small values of
$J$.
However, NBO analysis reveals a larger ligand-to-metal back charge
transfer for $J=0.7$~eV, which is consistent with the increased
occupancy of the Fe \dz orbital (Table~\ref{table}), and is expected
to cause variational energetic lowering at higher values of $J$ due to
electronic delocalization.

Compared to previously reported DFT studies, including our own DFT+$U$
study of the same system~\cite{cole}, the present study predicts a
significantly larger charge on the \oxy molecule ($-1.1$~e versus
$-0.5$~e).
This charge is expected to stabilize the \oxy molecule in the Mb
protein {\it via} hydrogen bond interactions with His\,64.
Hence, we propose that both dynamical and multi-reference quantum
effects, and large system sizes, must be accounted for in order to correctly
determine the energetics of ligand binding in proteins with strongly
correlated subspaces.

 \section{Conclusions}

We have presented the application of a newly-developed methodology,
designed to treat strong electronic interaction and multi-reference 
effects in systems of relatively very large numbers of atoms, 
to a molecule of important biological function.  
In particular, we have found that the Hund's coupling $J$
is the crucial ingredient necessary to increase 
the multi-reference high-spin character of the ground-state, and so to
bring the binding energetics into qualitative agreement with
experiment. This provides a route to the solution of
a long-standing problem in the 
density-functional theory based simulation of heme proteins, 
which often underestimates the Hund's coupling and 
incorrectly-describes multi-reference effects,
namely an unphysically large 
imbalance of CO and O$_2$ binding energies.
Our many-body description of the ligated myoglobin ground-state
is further supported by quantitative agreement with experimental 
findings on both the ligand-dependence of the optical absorption spectra
and the nature of the $\pi$-bonding in Fe--\oxyd.
Our approach, optimized to describe molecules and nano-particles involving 
transition metal ions, supports a large range of applications, e.g., to
strongly correlated oxide nano-particles \cite{spion_drug_delivery} 
or to enzymes \cite{enzyme_transition_metal}.


\section{Methodology}

In this work, we have carried out a detailed theoretical study
of the electronic structure of the myoglobin molecule by 
means of a  combination~\cite{our_paper_vo2,linearscalingdftu}
of linear-scaling density-functional theory (DFT) 
and the dynamical mean-field theory 
approximation (DFT+DMFT)~\cite{OLD_GABIS_REVIEW,
dca_reference_cluster_dmft},
a model which treats subspace-local dynamical, finite-temperature 
and multi-determinantal effects, for given Hamiltonian
parameters.

The ONETEP linear-scaling DFT code~\cite{onetep,linearscalingdftu,onetep_ref6} 
 was used to obtain the DFT ground-state.
The ONETEP method is particularly advanced in terms of its accuracy, 
equivalent to that of a plane-wave method,
which is arrived at by means of an \emph{in situ} 
variational optimization of the  
expansion coefficients of a minimal set of spatially-truncated 
Nonorthogonal Generalized Wannier 
Functions~\cite{onetep_ref_ngwf} (NGWFs), and 
is based on direct minimization of
the total-energy with respect to the single-particle density-matrix. 
The use of a minimal, optimized Wannier function 
representation of the density-matrix
allows for the DFT ground state to be solved 
with relative ease in large systems, 
particularly in molecules where their explicit truncation implies that the
addition of vacuum does not increase the computational cost.

Preparation of the structures for DFT+DMFT analysis have been described
in detail elsewhere~\cite{cole}.
Briefly, the computational models are based on the X-ray crystal
structures of sperm whale Mb in oxygenated and carbonmonoxygenated
ligation states (PDB: 1A6M, 1A6G)~\cite{vojtechovsky99}.
The heme group, ligand and 53 closest residues (1007 atoms in total)
were extracted from the Mb\oxy crystal structure and optimized using
spin-polarized DFT, with the PBE~\cite{PhysRevLett.77.3865} 
gradient-corrected exchange-correlation 
functional augmented by damped London potentials to
describe van der Waals interactions~\cite{hill09}.
Following optimization, the heme group and three closest residues were
replaced by their positions in the MbCO crystal structure and
re-optimized.
This scheme ensures that energy differences are directly attributable
to local changes in the binding site, while accounting for long-ranged
polarization and constraints of the protein scaffold.
The DFT binding energy was converged to better than $0.02$~eV with respect to 
changes in the plane-wave energy cutoff and NGWF cutoff radii, 
and no additional restrictions on the variational freedom, such as the density 
kernel truncation, were invoked.

We refined our DFT calculations using the DFT+DMFT 
method~\cite{OLD_GABIS_REVIEW,dca_reference_cluster_dmft} 
in order to 
obtain a more accurate treatment of strong electronic correlation effects.
In particular, DMFT introduces both quantum and thermal fluctuations,
which are multi-reference effects not captured at the level of the Kohn-Sham DFT.
In this, the Mb molecule was mapped, within DMFT, to an 
Anderson impurity model (AIM) Hamiltonian~\cite{heme_aim_kondo},
and we used a recently developed extended Lanczos solver~\cite{cpt_ed_solver} to obtain the DMFT self energy.
Since only a single impurity site ($3d$ orbital subspace) is present, the system becomes crystal momentum independent in the molecular limit, and since the Kohn-Sham Green's function is computed in full, by inversion, before projection onto the impurity subspace, the Anderson impurity mapping is effectively exact,  and the necessity of invoking the DMFT self-consistency is not required. However, in DFT+DMFT there is also a charge self-consistency cycle, albeit not routinely invoked at present due to computational cost, where the DFT+DMFT density kernel is used to generate a new Kohn-Sham Hamiltonian, which in turn provides a new input to the DMFT; the procedure being repeated until convergence is achieved. In this work, our data are obtained in the absence of charge self-consistency, however we checked that the corrections are small. Indeed, for Mb\oxy at $J=0.7$~eV, the changes obtained by converging the charge self-consistent DFT+DMFT induce a change in the energy of $\Delta E = -0.09$~eV, which corresponds to the energy of Mb\oxy at $J=0.72$~eV when the charge self-consistency is absent. Other changes are also small, for example, the chemical potential changes by $+0.023$~eV, and we find a change in the Fe charge of $+0.007$~e. All these variations are consistent with a renormalized $J$ ($J$ increased by ~3\% at $J=0.7$~eV).

To obtain the Kohn-Sham Green's function,  we performed the matrix inversion, as well as all 
matrix multiplications involved in the DMFT algorithm,
on graphical computational units (GPUs) using a tailor-made parallel implementation of the Cholesky decomposition 
written in the CUDA programming language.

Electronic correlation effects are described within the localized subspace by the 
Slater-Kanamori form of the Anderson impurity Hamiltonian~\cite{slater_kanamori_interaction,hund_coupling_kanamori},
specifically:
\begin{align}
\label{hint}
\mathcal{H}_{U} ={}& U \sum \limits_{m} {n_{m \up} n_{m \dn}} + \left(
U' -\frac{J}{2} \right)\sum\limits_{m>m'}{n_{m}n_{m'}}  \\
\nonumber {}&
-J \sum\limits_{m>m'}{\left( 2 \bold{S}_m \bold{S}_{m'} + \left(  d^\dagger_{m\up} d^\dagger_{m \dn } d_{m' \up} d_{m' \dn}   \right) \right)    },
\end{align}
where $m, m'$ are orbital indices, $d_{m\sigma}$ ($d^\dagger_{m\sigma}$) annihilates (creates) an electron with spin $\sigma$ in the orbital $m$,  
$n_m$ is the orbital occupation operator. 
The first term describes the effect of 
intra-orbital Coulomb repulsion, parametrized by $U$, and the second
term describes the inter-orbital repulsion, proportional to $U'$,
which is renormalized by the Hund's exchange coupling 
parameter $J$ in order
to ensure a fully rotationally invariant Hamiltonian (for 
further information on this topic, we
refer the reader to Ref.~\onlinecite{imada_mott_review}). 
The third term is the Hund's rule exchange coupling, 
described by a spin exchange coupling of amplitude $J$.
$\mathbf{S}_{m}$ denotes the spin corresponding to orbital $m$,
so that $S_{m}=\frac{1}{2}d_{m s}^{\dagger}
\vec{\sigma}_{s s'}
d_{m s' }$, where $\vec{\sigma}$ is the vector of Pauli matrices
indexed by $s$ and $s'$.
In this work, we used $U=4$~eV for the screened 
Coulomb interaction~\onlinecite{heme_marzari}, and
we explored the dependence of several observables on the Hund's coupling (in the range $J=0-1.5$~eV).
Our DMFT calculations were carried out at room 
temperature, $T=293$~K.
In this work, we used the canonical 
form of the double-counting potential $V_{\mathrm{dc}}$, 
given by:
\begin{equation}
\label{dc}
V_{\mathrm{dc}}= U^{\textrm{av}} \left( n_d - \frac{1}{2} \right) - 
\frac{J}{2} \left( n_d - 1 \right),
\end{equation}
assuming paramagnetic occupancy $n_d = 2 n_{d\sigma}$ 
of the $d$ orbitals.
Here, the parameter  $U^{\textrm{av}}$ 
is the intra- and inter-orbital averaged repulsion~\cite{hund_coupling_averaged_double_counting}.
In our calculations we found that the DMFT solution remains paramagnetic, although the possibility
of spontaneous formation of a magnetic moment (spin symmetry broken state) was allowed for. 
However, the low energy states are in a quantum
superposition of polarized states, giving a fluctuating magnetic moment at the iron site.
The theoretical optical absorption was obtained in DFT+DMFT 
within the linear-response regime (Kubo formalism),
in the \emphasize{no-vertex-corrections} approximation~\cite{millis_optical_conductivity_review}, 
where it is given by:
\begin{align}
 \sigma(\omega) ={}&  \frac{2 \pi e^2 \hbar }{\Omega }  \int  
 d \omega' \;
 \frac{  f( \omega'-\omega)-f \left( \omega' \right)}{\omega} \\
 {}&
 \times 
 \left( \bold{\rho^{\alpha \beta}} \left( \omega' - \omega \right)  \bold{v}_{\beta \gamma}  \bold{\rho^{\gamma \delta}} \left(\omega' \right) \cdot 
 \bold{v}_{\delta \alpha} \right), \nonumber
\end{align}
and the factor of two accounts for spin-degeneracy, 
$\Omega$ is the simulation-cell volume, $e$ is the electron charge,
$ \hbar $ is the reduced Planck constant, 
$f \left( \omega \right)$ is the Fermi-Dirac distribution, 
and $\bold{\rho}^{\alpha \beta}$ is the  density-matrix
given by the frequency-integral of the interacting DFT+DMFT Green's 
function.
The matrix elements of the velocity operator, $\bold{v}_{\alpha \beta}$,
noting that we do not invoke the Peierls substitution~\cite{millis_optical_conductivity_review},
are given by:
\begin{equation}
\bold{v}_{\alpha \beta}= 
- \frac{i \hbar }{ m_e }
\langle \phi_\alpha \rvert  \bold{\nabla} \lvert \phi_\beta \rangle
+ \frac{i}{\hbar} 
\langle \phi_\alpha \rvert  
\left[ \hat{V}_{nl}, \mathbf{r} \right] \lvert \phi_\beta \rangle.
\end{equation} 
This expression is general to the 
NGWF representation~\cite{optical_conductivity_non_orthogonal_basis},
used in this work,
where the contribution to the non-interacting  
Hamiltonian due to the non-local part of the 
norm-conserving pseudopotentials~\cite{PhysRevB.84.165131,
PhysRevB.44.13071},
represented by $\hat{V}_{nl}$, is included.
Once the self energy matrix is obtained, it can be used to correct the DFT total energy with
the DMFT correction \cite{dmft_dft_total_energy,dft_dmft_total_energy_charge_self_consistence}:
\begin{equation}
E = E_{DFT}\left[ \hat{\rho} \right] - \sum\limits_{\bold k \nu}{' \epsilon _{\bold k \nu}} + \text{Tr} \left[  \hat H_{DFT} \hat G \right]  + \langle  \hat H_U  \rangle - E_{DC} ,
\end{equation}
where $\hat H_U$ indicates the many body interaction vertex of the DMFT, 
and the primed sum is over the occupied states. 
The symbol ``Tr'' indicates the one-electron trace for a generic representation and the sum over the Matsubara frequencies $i\omega_n$ of the finite-temperature many-body formalism. The interaction term $\hat H_U$ is obtained with the Galitskii-Migdal formula \cite{total_energy_dft_dmft}:
\begin{equation}
\langle \hat H_U \rangle = \frac{1}{2} \text{Tr} \left[  \hat\Sigma \hat G   \right],
\end{equation} 
$\hat \Sigma$ ($\hat G$) is the self-energy (Green's function) matrix in the NGWF representation.  
We note that both the self energy and the Green's function are slowly decaying functions, hence the trace over Matsubara frequencies has to be done with care~\cite {dmft_dft_total_energy,dft_dmft_total_energy_charge_self_consistence}.
Finally, the double-counting correction $E_{DC}$ must be introduced, since the contribution of interactions between the correlated orbitals to the total energy is already partially included in the exchange-correlation potential derived from DFT.  
The most commonly used form of the double-counting term is~\cite{hund_coupling_averaged_double_counting}:
\begin{equation}
E_{\mathrm{dc}} = \frac{U^{\textrm{av}}}{2} n_d \left( n_d - 1 \right)  - \frac{J}{2} \sum\limits_\sigma{n_{d\sigma} \left( n_{d\sigma}-1 \right)}.
\end{equation} 

A new approach, developed in this work, is the generation 
of natural bond orbitals based 
on the many-body Green's function provided by DFT+DMFT, 
in order to obtain greater chemical insight into the ligand binding process.
Natural bond orbitals (NBOs)~\cite{reed88}
are post-processed linear-combinations of the 
basis functions in which 
the density-matrix is expanded, such
that the projection of the density-matrix
onto the subspace formed by atom-based 
and atom-pair based subsets of basis-functions is maximally diagonal. 
In the current calculations the basis-functions in question are 
NGWFs~\cite{onetep_ref_ngwf}, transformed to NBOs using
the \emph{NBO 5} programme~\cite{nbo5}, recently interfaced
to ONETEP, as described in Ref.~\cite{lee13}.
This procedure is carried out in such a manner that 
the final NBOs are then categorized into largely-occupied
bonding and lone-pair orbitals, 
and largely-vacant anti-bonding and Rydberg orbitals.
While normally applied to Kohn-Sham density-functional theory, to date, 
the NBO generation procedure is independent of the model 
(and so the Hamiltonian and self-energy) generating the density-matrix,
and so we may apply it to the density-matrix 
integrated from the DFT+DMFT full Green's function, for the first time. 
The resulting many-body NBOs largely retain the familiar profile 
of DFT-based NBOs, in this study, but their occupancies may be 
expected to deviate further from integer values due to quantum-mechanical
and finite-temperature multi-reference effects captured within DFT+DMFT.

We computed the energy of MbCO and MbO$_2$ as a function of the Hund's
exchange coupling $J$.
Defining
$\Delta E_{CO} = E_{MbCO} -E_{CO}$ and 
$\Delta E_{O_2} = E_{MbO_2} -E_{O_2}$, 
the binding energy difference $\Delta \Delta E$ is given by
$\Delta \Delta E = \Delta E_{O_2} - \Delta E_{CO}$. 
When $\Delta E_{CO}=\Delta E_{O_2}$, the MbO$_2$ and MbCO 
binding energies are identical.
In comparisons with the experimental relative free energy of binding,
we have assumed that the relative change in entropy of the two ligands
upon binding is zero, which is a reasonable approximation for two
sterically similar diatomic ligands.
%


\begin{acknowledgments}

We are grateful to Tanusri Saha-Dasgupta, Nicholas Hine, Gabriel Kotliar, Louis Lee, Peter Littlewood, and Andy Millis for helpful discussions.
DJC is supported by a Marie Curie International Outgoing Fellowship within the 7th 
European Community Framework Programme.
Calculations were performed on BlueGene/Q at the STFC Hartree Centre under project HCBG005 and 
on the Cambridge HPC Service, funded by EPSRC grants EP/J017639/1 and EP/F032773/1.
We gratefully acknowledge the support of NVIDIA Corporation with the donation of Tesla K20 GPU used for this research.
\end{acknowledgments}


\end{article}









\end{document}